\documentclass[10pt,twocolumn,aps,pra,showpacs]{revtex4}

\usepackage{bbm}

\begin{document}

\title{Faster quantum searching with almost arbitrary operators}         
\author{Avatar Tulsi\\
        {\small Department of Physics, IIT Bombay, Mumbai-400076, India}}  

\email{tulsi9@gmail.com}

\begin{abstract}

Grover's search algorithm drives a quantum system from an initial state $|s\rangle$ to a desired final state $|t\rangle$ by using selective phase inversions of these two states. In ~\cite{general}, we studied a generalization of Grover's algorithm which relaxes the assumption of the efficient implementation of $I_{s}$, the selective phase inversion of the initial state, also known as \emph{diffusion} operator. This assumption is known to become a serious handicap in cases of physical interest~\cite{spatial,kato,shenvi,realambainis}. Our general search algorithm works with almost arbitrary diffusion operator $D_{s}$ with only restriction of having $|s\rangle$ as one of its eigenstates. The price that we pay for using arbitrary operator is an increase in the number of oracle queries by a factor of $O(B)$, where $B$ is a characteristic of the eigenspectrum of $D_{s}$ and it can be large in some situations. Here we show that by using quantum fourier transform, we can regain the optimal query complexity of Grover's algorithm without losing the freedom of using arbitrary diffusion operators for quantum searching. However, the total number of operators required by algorithm is still $O(B)$ times more than that of Grover's algorithm. So our algorithm offers advantage only if oracle operator is computationally more expensive than diffusion operator, which is true in most search problems. 

\end{abstract}

\pacs{03.67.Ac}

\maketitle

\section{INTRODUCTION}

Suppose we have a quantum computer initially in a \emph{source} state $|s\rangle$ and we want to evolve it to a final state $|t\rangle$, say the \emph{target} state. One way to do this is famous Grover's algorithm~\cite{grover,qaa1,qaa2} which assumes that we can efficiently implement the selective phase inversion operators, $I_{s}$ and $I_{t}$, of these two states. Grover's algorithm keeps on iterating the Grover's search operator $\mathcal{G} = I_{s}I_{t}$ on the source state $|s\rangle$ to get the target state $|t\rangle$. The required number of iterations is $\pi/4\alpha$, where $\alpha = |\langle t|s\rangle|$. 

While using Grover's algorithm for the search problem, we choose $|s\rangle$ to be the uniform superposition of all $N$ basis states to be searched i.e. $|s\rangle = \sum_{i}|i\rangle/\sqrt{N}$. In case of a unique solution to the search problem, the target state $|t\rangle$ is a unique basis state and $\alpha = |\langle t|s\rangle| = 1/\sqrt{N}$. Thus Grover's algorithm outputs a solution in just $O(\sqrt{N})$ time steps which is quadratically faster than \emph{classical} search algorithms taking $O(N)$ time steps. 

Grover's algorithm is proved to be strictly optimal~\cite{optimal} and the assumption of efficient implementation of $I_{s}$ and $I_{t}$ is justified in many situations. The oracle operator $I_{t}$ always needs an oracle query for implementation whereas implementation of the diffusion operator $I_{s}$ is dictated by physical constraints and sometimes it may become a serious problem. For example, in the case of two-dimensional spatial search~\cite{spatial}, each implementation of $I_{s}$ takes $\sqrt{N}$ time steps and hence the total time complexity of Grover's algorithm becomes $\sqrt{N} \times \sqrt{N} = N$ time steps, which is no better than classical algorithms. 

Earlier, several attempts have been done to study quantum search algorithms with more general diffusion operators in place of $I_{s}$. For example, Kato studied the case when $I_{s}$ is replaced by an operator made up of only single qubit gates~\cite{kato}. Ambainis studied the case when $I_{s}$ is replaced by a real operator with $|s\rangle$ as its eigenstate~\cite{realambainis}. In ~\cite{general}, we presented a unified framework for all such generalizations and we studied the case when $I_{s}$ is replaced by any arbitrary operator $D_{s}$ with the only restriction of having $|s\rangle$ as one of its eigenstates. This restriction seems to be more or less \emph{justified} as the search operator should have some special connection with the source state. 

To be more concrete, let the normalized eigenspectrum of $D_{s}$ be given by $D_{s}|\ell\rangle = e^{\imath\theta_{\ell}}|\ell\rangle$ with $|\ell\rangle$ as the eigenstates and $e^{\imath\theta_{\ell}}$ ($\theta_{\ell}$) as the corresponding eigenvalues (eigenphases). Since a global phase is irrelevant in quantum dynamics, we choose $D_{s}|s\rangle = |s\rangle$, i.e. $\theta_{\ell=s} = 0$. In ~\cite{general}, we studied the iteration of general search operator $\mathcal{S} = D_{s}I_{t}$ on $|s\rangle$ by analyzing the eigenspectrum of $\mathcal{S}$. We found that the performance of quantum search algorithm depends upon the quantities $\Lambda_{1}$ and $\Lambda_{2}$, where
\begin{equation}
\Lambda_{p} = \sum_{\ell \neq s}|\langle \ell|t\rangle|^{2}\cot^{p}\frac{\theta_{\ell}}{2} \label{momentdefine}
\end{equation}
is the $p^{\rm th}$ moment of $\cot\frac{\theta_{\ell}}{2}$ with respect to the distribution $|\langle \ell |t\rangle|^{2}$ over all $\ell \neq s$. 

We found that $\Lambda{1}$ should be very close to zero for a successful quantum search. We also presented an algorithm in Section IV.A of ~\cite{general} which uses an ancilla qubit to control the applications of $D_{s}$ and $D_{s}^{\dagger}$ in a clever way so that effectively $\Lambda_{1}$ becomes zero. In this paper, we restrict ourselves to the case $\Lambda_{1} = 0$. In case, it is not so, we can always use just-mentioned algorithm to make it so. If $\Lambda_{1} = 0$, we get the target state using $(\pi/4\alpha)B^{3}$ oracle queries where $B = \sqrt{1+\Lambda_{2}}$. In Section IV.B of ~\cite{general}, it was shown that by controlling the operators using an ancilla qubit, we can get a faster algorithm which uses only $(3\sqrt{3}/2)(\pi/4\alpha)B$ oracle queries. But even now, this algorithm is slow compared to the optimal Grover's algorithm by a factor of $O(B)$. 

In this paper, essentially, we show that by using quantum fourier transform, we can manipulate the operation of $D_{s}$ cleverly to design a new operator for which $B = O(1)$. Thus we can achieve the optimal performance of Grover's algorithm with almost any operator $D_{s}$. In next section, we present a brief review of the results of our analysis of general quantum search algorithm presented in ~\cite{general}. In Section III, we present our new algorithm with optimal performance. Then we conclude in Section IV.

\section{GENERAL QUANTUM SEARCH: A BRIEF REVIEW}

Here we briefly discuss the dynamics of general quantum search algorithm presented in ~\cite{general}. We skip the details for which readers are referred to the original paper. This algorithm iterates the operator $\mathcal{S} = D_{s}I_{t}^{\phi}$ on $|s\rangle$ to take it close to $|t\rangle$. Here $D_{s}$ is as defined earlier and $I_{t}^{\phi}$ is the selective phase rotation of target state by angle of $\phi$. Without any loss of generality, we take $\phi = \pi$ so that $I_{t}^{\phi}$ is the selective phase inversion $I_{t}$ of the target state. Also, we assume $|s\rangle$ to be a non-degenerate eigenstate for simplicity. 
Let the normalized eigenspectrum of $D_{s}$ be given by $D_{s}|\ell\rangle = e^{\imath\theta_{\ell}}|\ell\rangle$. By convention, $\theta_{\ell = s} = 0$. Let other eigenvalues satisfy
\begin{equation}
|\theta_{\ell \neq s}| \geq \theta_{\rm min} > 0,\ \ \theta_{\ell} \in [-\pi,\pi] \label{othereigenvalues}
\end{equation} 
To study the iteration of $\mathcal{S}$ on $|s\rangle$, we need to find its eigenspectrum. We found that only two eigenstates $|\lambda_{\pm}\rangle$ with the corresponding eigenvalues $e^{\imath \lambda_{\pm}}$ of $\mathcal{S}$ are relevant for our algorithm. Under the assumption, $|\lambda_{\pm}| \ll \theta_{\rm min}$, the initial state $|s\rangle$ is almost completely spanned by two eigenstates $|\lambda_{\pm}\rangle$. 
The eigenvalues $\lambda_{\pm}$ are given by 
\begin{equation}
\lambda_{\pm} = \pm\frac{2\alpha}{B}(\tan \eta)^{\pm 1}\ \ \cot 2\eta = \frac{\Lambda_{1}}{2\alpha B}\ .\label{solutions}
\end{equation}
where
\begin{equation}
B = \sqrt{1 + \Lambda_{2}}\ ,\ \Lambda_{p} = \sum_{\ell \neq s}|\langle \ell|t\rangle|^{2}\cot^{p}\frac{\theta_{\ell}}{2}\ . \label{definitionLambda}
\end{equation}
As $\alpha = |\langle s|t\rangle| \ll 1$, $\sum_{\ell \neq s}|\langle \ell |t\rangle|^{2}$ is very close to $1$ and hence above equation gives
\begin{equation}
B^{2} = \sum_{\ell \neq s} |\langle \ell|t\rangle|^{2}\left(1+\cot^{2}\frac{\theta_{\ell}}{2}\right)= \sum_{\ell \neq s} |\langle \ell|t\rangle|^{2}\frac{1}{\sin^{2}\frac{\theta_{\ell}}{2}}\ .\label{Bexpr}
\end{equation}
We only consider the case when $\Lambda_{1} = 0$ as only then a successful quantum search is possible. If $\Lambda_{1}\neq 0$, we can always use the algorithm presented in Section IV.A to design a new operator using $D_{s}$ and $D_{s}^{\dagger}$ for which $\Lambda_{1} = 0$. In this case, \ref{solutions} indicates that 
\begin{equation}
\Lambda_{1} = 0 \Longrightarrow \eta = \frac{\pi}{4},\ \lambda_{\pm} = \pm\frac{2\alpha}{B}\ .
\end{equation} 

With $\eta = \pi/4$ and $\phi = \pi$, Eq. (23) and (24) of ~\cite{general} gives us the initial state $|s\rangle$ and the effect of iterating $\mathcal{S}$ on $|s\rangle$ in terms of two relevant eigenstates. We have
\begin{equation}
|s\rangle = -\imath/\sqrt{2}[e^{\imath \lambda_{+}/2}|\lambda_{+}\rangle - e^{\imath \lambda_{-}/2}|\lambda_{-}\rangle], \label{slambdapmexpansion}
\end{equation}
and
\begin{equation}
\mathcal{S}^{q}|s\rangle =  -\imath/\sqrt{2} [e^{\imath q'\lambda_{+}}|\lambda_{+}\rangle - e^{\imath q'\lambda_{-}}|\lambda_{-}\rangle],
 \label{stateexpand}
\end{equation}
where $q'  = q+\frac{1}{2}$.

For $q = q_{\rm m} \approx \pi/2|\lambda_{\pm}| = \pi B/4\alpha$, the state $\mathcal{S}^{q_{\rm m}}|s\rangle$ is very close to the state $|w\rangle$ given by 
\begin{equation}
\mathcal{S}^{q_{\rm m}}|s\rangle = |w\rangle = 1/\sqrt{2}(|\lambda_{+}\rangle + |\lambda_{-}\rangle).
\end{equation}
As shown in ~\cite{general}, we have $|\langle t|w\rangle| = 1/B$.
So we get the target state with a probability of $1/B^{2}$ after $\pi B/4\alpha$ iterations of $\mathcal{S}$ on the initial state $|s\rangle$. As each application of $\mathcal{S}$ needs one query to implement $I_{t}$, we need a total of $(\pi/4\alpha)B^{3}$ oracle queries to get the target state. Section IV.B of ~\cite{general} presents an algorithm which improves the query complexity to $O(\pi B/4\alpha)$ by controlling the applications of $D_{s}$ using an ancilla qubit. 

However, we still need $O(B)$ times more queries compared to the optimal Grover's algorithm. We can see from \ref{definitionLambda} that $\Lambda_{2} \leq 1/\theta_{\rm min}^{2}$ so for $\theta_{\rm min} \ll 1$, the only upper bound that we have on $B$ is $1/\theta_{\rm min}$, which can in general be arbitrarily large. Thus we need some trick to reduce the value of $B$. One way is to use quantum fourier transform to reliably distinguish the eigenstates of $D_{s}$ using Phase Estimation algorithm. As $\theta_{\ell = s} = 0$ and $|\theta_{\ell \neq s}| \geq \theta_{\rm min}$, we need $O(1/\theta_{\rm min})$ applications of $D_{s}$ to achieve such a reliable distinction after which, we can selectively invert the phase of $|s\rangle$ state to implement $I_{s}$. Such a scheme will take $O(1/\alpha \theta_{\rm min})$ applications of $D_{s}$ and $O(1/\alpha)$ oracle queries. 

We point out that $1/\theta_{\rm min}$ is just an upper bound for $B$ hence above-mentioned simple scheme may not be more efficient than general quantum search algorithm with time complexity $O(B/\alpha)$. For example, in case of two-dimensional spatial search, $1/\theta_{\rm min} = O(\sqrt{N})$ but $B = O(\sqrt{\ln N})$ so $B$ is much smaller than $1/\theta_{\rm min}$. In next section, we show that the quantum fourier transform can be used in a more clever way to get a successful quantum searching using only $O(B/\alpha)$ applications of $D_{s}$ and $O(1/\alpha)$ oracle queries.

\section{FASTER ALGORITHM}

The new algorithm basically reduces the effective value of $B$ to $O(1)$. To get the basic idea, first consider a simple case when the operator $D_{s}$ is iterated $r$ times to get a new operator $D_{s}^{r}$ with its eigenspectrum given by $D_{s}^{r}|\ell\rangle = e^{\imath r\theta_{\ell}}|\ell\rangle$. As given by \ref{Bexpr}, the corresponding value of $(B_{r})^{2}$ is
\begin{equation}
(B_{r})^{2} = \sum_{\ell \neq s} \frac{|\langle \ell |t\rangle|^{2}}{\sin^{2}\frac{r\theta_{\ell}}{2}}\ .
\end{equation}
As long as $\theta_{\ell}/2 \ll 1/r$, we have $\sin(r\theta_{\ell}/2) \approx r\sin(\theta_{\ell}/2)$ and so $B_{r} \approx B/r$. Thus $B_{r}$ can be made $O(1)$ by choosing $r = O(B)$. But this simple scheme works only if $\theta_{\ell}/2 \ll 1/r$ for all $\ell$, not in the general case. In fact, we can see that if $\theta_{\ell}$ is sufficiently close to $2n\pi/r$ for any integer $n$, then $B_{r}$ will diverge.

We now show that the quantum fourier transform can be used in more general case to make $B = O(1)$. Basically we achieve some level of distinction among the eigenstates of $D_{s}$ and then to effectively nullify the contribution of those eigenstates for which $\theta_{\ell}/2$ is not much smaller than $1/r$.

Let $\mathcal{H}_{N}$ denote the Hilbert space of our main quantum system which we want to evolve from $|s\rangle$ to $|t\rangle$ state. We attach an ancilla quantum system of $m$ qubits to our main system and let $\mathcal{H}_{M}$ ($M = 2^{m}$) denote the corresponding Hilbert space with its basis states $|j\rangle$, $j\in \{0,1,\ldots,2^{m}-1\}$. This ancilla system will be used for quantum fourier transform as done in Phase Estimation algorithm. We work in the joint Hilbert space $\mathcal{H} = \mathcal{H}_{M} \otimes \mathcal{H}_{N}$.

\subsection{Phase Estimation Algorithm}

We first consider the operator $\mathcal{P}$ corresponding to the phase estimation algorithm. Let the initial state of our main quantum system be $|\ell\rangle$, an eigenstate of $D_{s}$, and let the initial state of our ancilla quantum system be $|\hat{0}\rangle$ in which all $m$ qubits are in $|0\rangle$ state. So the initial state for PEA is
\begin{equation}
|\hat{0},\ell\rangle = |\hat{0}\rangle_{M} \otimes |\ell\rangle_{N},
\end{equation}
where the subscripts $M$ and $N$ denote the corresponding Hilbert space of quantum states. We omit these subscripts for simplicity.

The phase estimation algorithm operator $\mathcal{P}$ is a successive application of three operators on the state $|\hat{0},\ell\rangle$, i.e.
\begin{equation}
\mathcal{P} = (\mathcal{F}\otimes \mathbbm{1}_{N})(c_{j}D_{s}^{j})(W\otimes \mathbbm{1}_{N})|\hat{0},\ell\rangle.
\end{equation} 

The first operator $\mathcal{P}_{1} = W\otimes \mathbbm{1}_{N}$ applies Walsh-Hadamard transform on the ancilla system and leaves the main system unchanged. Thus the ancilla system is transformed to an uniform superposition of all $m$ basis states and we get
\begin{equation}
\mathcal{P}_{1}|\hat{0},\ell\rangle = \frac{1}{2^{m/2}}\sum_{j = 0}^{2^{m}-1}|j\rangle|\ell\rangle.
\end{equation}

The second operator $\mathcal{P}_{2} = c_{j}D_{s}^{j}$ is a controlled application of $D_{s}^{j}$ operator on the main quantum system, i.e. it applies $j$ iterations of $D_{s}$ on the main quantum system if and only if the ancilla quantum system is in the $|j\rangle$ state. \begin{equation}
\mathcal{P}_{2}\mathcal{P}_{1}|\hat{0},\ell\rangle = \frac{1}{2^{m/2}}\sum_{j = 0}^{2^{m}-1}|j\rangle D_{s}^{j}|\ell\rangle. 
\end{equation}
As the main quantum system is in an eigenstate $|\ell\rangle$ of $D_{s}$, we get
\begin{equation}
\mathcal{P}_{2}\mathcal{P}_{1}|\hat{0},\ell\rangle = \frac{1}{2^{m/2}}\sum_{j = 0}^{2^{m}-1}e^{\imath j\theta_{\ell}}|j\rangle |\ell\rangle. \label{P2state}
\end{equation}

The third operator $\mathcal{P}_{3} = \mathcal{F} \otimes \mathbbm{1}_{N}$ applies a quantum fourier transform on the ancilla quantum system but leaves the main system unchanged. The action of $\mathcal{F}$ in a $2^{m}$-dimensional Hilbert space on each basis state $|j\rangle$ is given by
\begin{equation}
\mathcal{F}|j\rangle = \frac{1}{2^{m/2}}\sum_{k =0}^{2^{m}-1}exp\left(2\pi \imath kj/2^{m}\right) |j\rangle. \label{QFTequation}
\end{equation}
where $k$ is the numerical value of the binary number represented by the bit string encoded by $|k\rangle$ state. For example, for $|k\rangle = |\hat{0}\rangle$, all qubits are in $|0\rangle$ state and hence $k = 0$.

As suggested by \ref{P2state}, before applying $\mathcal{F}$, the ancilla quantum system is in the state $(1/2^{m/2})\sum_{j}e^{\imath j\theta_{\ell}}|j\rangle$. So, using \ref{QFTequation}, we get
\begin{equation}
\mathcal{P}|\hat{0},\ell\rangle = |\theta_{\ell}\rangle|\ell\rangle,
\end{equation} 
where
\begin{equation}
|\theta_{\ell}\rangle = \frac{1}{2^{m}}\sum_{k,j=0}^{2^{m}-1}exp[\imath (2\pi jk/2^{m} – j\theta_{\ell})]|k\rangle.
\end{equation}
This is the standard output state of Phase Estimation algorithm and has been analyzed quite well in literature (for example, see Sec. 5.2.1 of ~\cite{phase}). We note that
\begin{equation}
\langle k|\theta_{\ell}\rangle = \frac{1}{2^{m}}\sum_{j=0}^{2^{m}-1}(exp[\imath(2\pi k/2^{m} -\theta_{\ell})])^{j},
\end{equation}
which is the sum of a geometric series and after little calculation, we get
\begin{equation}
|\langle k|\theta_{\ell}\rangle| = \frac{1}{2^{m}}\frac{\sin[\pi k – 2^{m-1}\theta_{\ell}]}{\sin[(\pi k – 2^{m-1}\theta_{\ell})/2^{m}]}\ .
\end{equation}

For the purpose of our algorithm, what matters is the overlap of $|\theta_{\ell}\rangle$ state with $|k = \hat{0}\rangle$ state for which $k=0$ and above equation gives us
\begin{equation}
|\langle \hat{0}|\theta_{\ell}\rangle| = \frac{1}{2^{m}}\frac{\sin[2^{m-1}\theta_{\ell}]}{\sin[\theta_{\ell}/2]}\ . \label{amplimod}  
\end{equation}
For $\ell =s$, $\theta_{\ell} = 0$ and in the limit $\theta_{\ell} \rightarrow 0$, the R.H.S. of above equation becomes $1$ so we have $|\theta_{s}\rangle = |\hat{0}\rangle$. So if the main quantum system is in $|s\rangle$ state, then the operator $\mathcal{P}$ leaves the entire quantum system unchanged.

\subsection{New operator}

The new operator that we design using $D_{s}$ as basic operator is given by $\mathcal{D} = \mathcal{P} \mathcal{C} \mathcal{P}^{\dagger}$, where $\mathcal{C}$ is a diagonal operator given by
\begin{equation}
\mathcal{C} = (c_{j\neq \hat{0}}(-\mathbbm{1}_{N}))(c_{\hat{0}}D_{s}^{r}). 
\end{equation}
First, $\mathcal{C}$ applies the operator $c_{\hat{0}}D_{s}^{r}$ which performs $r$ iterations of the operator $D_{s}$ on the main quantum system, if and only if the ancilla quantum system is in $|\hat{0}\rangle$ state. Then $\mathcal{C}$ applies the operator $c_{j\neq \hat{0}}(-\mathbbm{1}_{N})$ which inverts the phase of all basis states $|j\rangle$ of the ancilla quantum system except the $|j= \hat{0}\rangle$ state.
It is easy to check that the eigenstates and the corresponding eigenvalues of the operator $\mathcal{C}$ are given by
\begin{eqnarray}
|j =\hat{0}\rangle|\ell\rangle &\ ;\ & exp[\imath r\theta_{\ell}] \nonumber \\ 
|j\neq \hat{0}\rangle|\ell\rangle &\ ;\ & exp[\imath \pi] \ . \label{Ceigen}
\end{eqnarray}
As $\mathcal{D} = \mathcal{P} \mathcal{C} \mathcal{P}^{\dagger}$, we find that the eigenstates $|\ell'\rangle$ and the corresponding eigenvalues $e^{\imath \theta_{\ell}'}$ of our new operator $\mathcal{D}$ are given by
\begin{eqnarray}
\mathcal{P}(|j =\hat{0}\rangle|\ell\rangle) &\ ;\ & exp[\imath r\theta_{\ell}] \\ \nonumber
\mathcal{P}(|j\neq \hat{0}\rangle|\ell\rangle) &\ ;\ & exp[\imath \pi] \ . \label{Ceigen}
\end{eqnarray}
The eigenstate corresponding to $\theta_{\ell}' = 0$ is the effective source state $|s'\rangle$ and above equation gives $|s'\rangle = \mathcal{P}(|\hat{0}\rangle|s\rangle)$. As discussed in the previous subsection, when the main quantum system is in $|s\rangle$ state, then the operator $\mathcal{P}$ leaves the entire quantum system unchanged. So we have $|s'\rangle = |\hat{0}\rangle|s\rangle$.

Now consider the oracle operator used by our algorithm. In place of simply applying the selective phase inversion $I_{t}$ on the main quantum system, our new algorithm applies $I_{t}$ on the main system if and only if the ancilla system is in $|\hat{0}\rangle$ state. By doing this, basically we apply $I_{\hat{0},t}$, the selective phase inversion of $|\hat{0}\rangle|t\rangle$ state and our effective target state is $|t'\rangle = |\hat{0}\rangle|t\rangle$. 

With all these details of new operators, the value of $B'$ for $\mathcal{D}$ can be found. We use the expression \ref{Bexpr}, i.e.  $(B')^{2} = \sum_{\ell' \neq s'}(|\langle \ell'|t'\rangle|/\sin\frac{\theta_{\ell}'}{2})^{2}$. We separate this sum into two parts: $\Sigma_{1}$ and $\Sigma_{2}$ where $\Sigma_{1}$ is due to the eigenstates $\mathcal{P}(|j\neq \hat{0}\rangle|\ell\rangle)$ having eigenvalue $e^{\imath \pi}$, i.e. $\theta_{\ell}' = \pi$. As $\sin^{2}\frac{\theta_{\ell}'}{2} = 1$ for all such eigenstates and the sum $\sum_{\ell'} |\langle \ell'|t\rangle|^{2}$ is always less than $1$ for any set of $|\ell'\rangle$, so we see that $\Sigma_{1} \leq 1$.
 
To find $\Sigma_{2}$, we note that for remaining eigenstates, the eigenvalues are $e^{\imath r\theta_{\ell}}$ and the values of $|\langle \ell' |t'\rangle|$ are given by 
\begin{displaymath}
|\langle\hat{0}|\langle t|\mathcal{P}(|\hat{0}\rangle|\ell\rangle)
\end{displaymath}
which, using \ref{amplimod}, is equal to
\begin{displaymath}
|\langle t|\ell\rangle||\langle \hat{0}|\theta_{\ell}\rangle = \frac{1}{2^{m}}\frac{\sin[2^{m-1}\theta_{\ell}]}{\sin[\theta_{\ell}/2]}|\langle t|\ell\rangle| \ .
\end{displaymath} 
So the corresponding contribution to $(B')^{2}$ is
\begin{equation}
\sum_{\ell' \neq s'}\frac{|\langle \ell'|t'\rangle|^{2}}{\sin^{2}\frac{\theta_{\ell}'}{2}} = \frac{1}{2^{2m}}\sum_{\ell \neq s}\frac{|\langle \ell|t\rangle|^{2}}{\sin^{2}\frac{r\theta_{\ell}}{2}}\frac{\sin^{2}[2^{m-1}\theta_{\ell}]}{\sin^{2}\frac{\theta_{\ell}}{2}}  \label{tempeq}
\end{equation}
We choose
\begin{equation}
r = 2^{m}\ .
\end{equation}
Then it is easy to check that the numerator of one term in \ref{tempeq} exactly cancels the denominator of another term. It is this cancellation which effectively nullifies the contribution of those eigenstates which otherwise had a potential to make $B$ diverge as we discussed earlier. And as we have shown that such a cancellation can be achieved easily through quantum fourier transform. After this cancellation, we get
\begin{equation}
\Sigma_{2} = \frac{1}{2^{2m}}\sum_{\ell \neq s}\frac{|\langle\ell|t\rangle|^{2}}{\sin^{2}\frac{\theta_{\ell}}{2}} = \frac{B^{2}}{2^{2m}}\ .
\end{equation}
And so, we get
\begin{equation}
(B')^{2} = \Sigma_{1}+\Sigma_{2} \leq 1+\frac{B^{2}}{2^{2m}}\ .
\end{equation}
Thus, by choosing $m = \log_{2}B$, we can make $B' = \sqrt{2} = O(1)$ and we can get the target state using only $O(\pi/4\alpha)$ oracle queries, which is within a constant factor of the optimal performance by Grover's algorithm. 

However, we note that implementation of the new operator $\mathcal{D}$ requires $3B$ applications of the basic operator $D_{s}$, one each for $\mathcal{P}$, $\mathcal{P}^{\dagger}$, and $\mathcal{C}$. And as we need $O(\pi/4\alpha)$ iterations of the search operator, the total applications of $D_{s}$ required by algorithm is still $O(B/\alpha)$ as required by the original algorithm presented in ~\cite{general}. The improvement mainly comes because of the reduction in number of oracle queries which involves major computational resources in typical search problems.

\subsection{Algorithm}

Above analysis suggests the following algorithm \\ (1)We attach an ancilla quantum system of $m = \log_{2}B$ qubits to our main quantum system. \\ (2)We choose the initial state of all qubits of our ancilla system to be in $|0\rangle$ state. We choose the initial state of our main quantum system to be the source state $|s\rangle$. \\ (3)We perform $O(1/\alpha)$ iterations of the operator $\mathcal{D}I_{\hat{0},t}$ on the initial state. Note that $I_{\hat{0},t}$ is a controlled application of $I_{t}$ and needs one oracle query for implementation. Also, $\mathcal{D}$ needs $O(B)$ applications of $D_{s}$ for implementation. \\ (4)We measure the main system, which will be in the target state with probability close to $1$.

\section{DISCUSSION AND CONCLUSION}

We have presented an algorithm to achieve the optimal performance of Grover's algorithm in the case of more general diffusion operators $D_{s}$ used in place of $I_{s}$. A simple scheme of successive iterations of $D_{s}$ may not help in general cases as some \emph{bad} eigenstates of $D_{s}$ may cause the parameter $B$ to diverge. What we have basically shown is that Quantum Fourier Transform allows us to nullify the effect of these bad eigenstates.

In the example of two-dimensional spatial search $\alpha = 1/\sqrt{N}$ and $B = O(\sqrt{\ln N})$. Hence our algorithm can get the target state using only $O(\sqrt{N})$ oracle queries in contrast to the $O(\sqrt{N\ln N})$ performance of earlier known algorithms, see for example~\cite{faster}. However, our algorithm will need $O(\sqrt{N\ln N})$ applications of the local operator and as each local operator needs one time step for implementation, the total time complexity is still $O(\sqrt{N\ln N})$ time steps. But in the cases when oracle query becomes more expensive than implementing local operators, our algorithm offers an advantage.

Our algorithm offers a general framework to achieve the optimal oracle query performance in general cases. It shows that by using general diffusion operators, we don't need to compromise with the query complexity of quantum search algorithms. We believe that our algorithm can find important applications in search problems.

\end{document}